\documentclass[aps,prb,twocolumn,showpacs,10pt]{revtex4-1}
\usepackage{graphicx,amsmath,amssymb,amsfonts,sidecap,color}

\newcommand{\ket}[1]{\ensuremath{\left|#1\right\rangle}}

\makeatletter
\newcommand*{\rom}[1]{\expandafter\@slowromancap\romannumeral #1@}
\makeatother

\begin{document}

\title{Time-Reversal Invariant Parafermions in Interacting Rashba Nanowires}

\author{Jelena Klinovaja$^{1}$}
\author{Daniel Loss$^2$}
\affiliation{$^1$Department of Physics, Harvard University,  Cambridge, Massachusetts 02138, USA,}
\affiliation{$^2$Department of Physics, University of Basel,
             Klingelbergstrasse 82, CH-4056 Basel, Switzerland}
             
\date{\today}

\begin{abstract}
We propose a scheme to generate pairs of time-reversal invariant parafermions. Our setup consists of two quantum wires with Rashba spin orbit interactions coupled to an $s$-wave superconductor, in the presence of electron-electron interactions. The zero-energy bound states localized at the wire ends arise from the interplay between two types of proximity induced superconductivity: the usual intrawire superconductivity and the interwire superconductivity due to crossed Andreev reflections. If the latter dominates, which is the case for strong electron-electron interactions, the system supports Kramers pair of parafermions. Moreover, the scheme can  be extended to a two-dimensional sea of  time-reversal invariant parafermions.
\end{abstract}

\maketitle

\section {Introduction}

Topological properties of  condensed matter systems have attracted wide attention in recent years. In particular, localized bound states emerging at the interface between different topological regions have been studied intensely both theoretically and experimentally. Majorana fermions (MFs), zero-energy bound states with non-Abelian braid statistics, were predicted in several systems such as fractional quantum Hall effect (FQHE) systems,\cite{Read_2000} topological insulators, \cite{fu, Nagaosa_2009,Ando} optical lattices, \cite{Sato,demler_2011}  $p$-wave superconductors, \cite{potter_majoranas_2011} nanowires with  Rashba spin orbit interaction (SOI), \cite{lutchyn_majorana_wire_2010, oreg_majorana_wire_2010, alicea_majoranas_2010,mourik_signatures_2012,das_evidence_2012,deng_observation_2012, marcus_MF,Rokhinson,Goldhaber,Rotating_field} self-tuning RKKY systems, \cite{RKKY_Basel,RKKY_Simon,RKKY_Franz} and  graphene-like systems.\cite{Klinovaja_CNT,bilayer_MF_2012, MF_nanoribbon, 
MF_MOS_Zigzag, MF_MOS, MF_Bena}

Though MFs possess non-Abelian statistics, it is of  Ising type which is not sufficient for universal quantum computation, 
in contrast to Fibonacci anyons.\cite{Pachos_book}
The basic building blocks for the latter anyons are parafermions (PFs), also referred to as fractional MFs, which allow for more universal quantum operations than 
MFs.\cite{Fradkin_PF_1980,topology_barkeshli,Fendley_PF_2012,PF_Linder,Vaezi,
Ady_FMF,PF_Clarke,PF_Mong,bilayer_PFs,vaezi_2,PFs_Loss} 
Similarly to MFs, PFs are bound states that arise at the interface between two distinct topological phases. In contrast to MFs, however, PFs owe their peculiar properties to strong electron-electron interactions.
 As a result, most proposals to host PFs invoke edge states of FQHE systems, and 
to stabilize them at zero energy one relies on particle-hole symmetry generated by proximity to a superconductor. \cite{PF_Linder,Vaezi,Ady_FMF,PF_Clarke,PF_Mong,vaezi_2}
However, while strong magnetic fields are required for the FQHE, they are detrimental for 
superconductivity,  making  the experimental realization of such proposals challenging. \cite{PF_Mong,Braunecker} This has motivated us to search for alternatives to generate PFs with superconductivity but without magnetic fields. Indeed, we will show that by taking advantage of time-reversal invariance it is possible to construct
Kramers pairs of PFs, which can be considered as generalization of Kramers pairs of MFs studied before. \cite{TRI1,TRI2,TRI3,TRI4, TRI5,TRI6,TRI7, TRI8,TRI9,TRI10}   We are also motivated to work with one-dimensional systems where recent experiments have demonstrated proximity-induced superconductivity of  crossed Andreev type, \cite{Feinberg_2000,Schonenberger,Heilblum} strong electron-electron interaction, \cite{CNT_LL_exp,Amir_exp_1,Amir_exp_2,Dominik_exp_K} and high tunability of the chemical potential. \cite{mourik_signatures_2012,das_evidence_2012,deng_observation_2012, marcus_MF,Rokhinson,Goldhaber} Moreover, the class of materials suitable for our scheme is larger than  for schemes with magnetic field since we do not require large $g$-factors.

\begin{figure}[t!]
\includegraphics[width=0.8\linewidth]{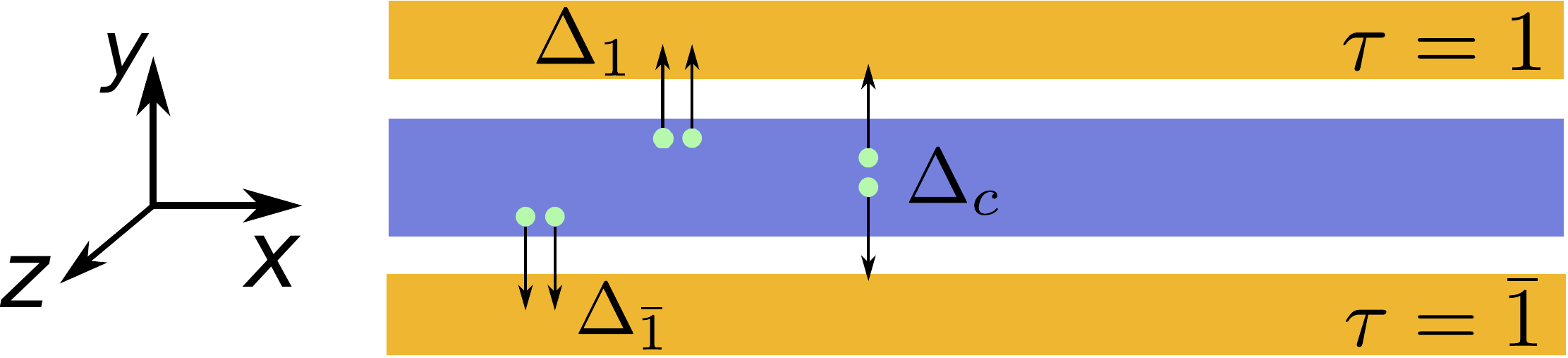}
\caption{Sketch of two Rashba QWs (yellow strips) coupled to an $s$-wave superconductor (blue strip). The SOI field points in positive (negative) direction, say, along the $z$ azis for the upper (lower) QW, $\tau=1$ ($\tau=\bar 1$). The intrawire proximity induced superconductivity of strength $\Delta_\tau$ corresponds to a Cooper pair (pair of green dots) tunneling as a whole into the $\tau$-wire. The interwire proximity induced superconductivity of strength $\Delta_c$ corresponds to crossed Andreev reflection into both QWs, which dominates for strong electron-electron interaction assumed here. }
\label{bundle}
\end{figure}

The setup we consider (see Fig. \ref{bundle}) consists of two one-dimensional channels, or quantum wires (QWs) with the Rashba SOI. The QWs are close to an $s$-wave superconductor resulting in proximity induced superconductivity. In general, there are two types of pairing terms. The first one is intrawire  pairing corresponding to tunneling of  Cooper pairs as a whole to either of the QWs. The second type is the interwire  pairing corresponding to  `crossed Andreev reflection' \cite{Feinberg_2000} where the Cooper pair gets split into two different channels. Such processes dominate in the regime of strong electron-electron interactions. \cite{Two_LL_SC_Patrik,Andreev_Bena,Andreev_Yaroslav}
In this case, the system is in the topological phase with bound states localized at the system ends. If the chemical potential is tuned close 
to the SOI energy, the system supports two MFs at each end that are time-reversal partners of each other. \cite{TRI8,TRI5} More strikingly, if the chemical potential is lowered, {\it e.g.} to one nineth of the SOI energy, and electron-electron interactions are strong, the zero-energy ground state contains three PF Kramers pairs.  However, similar to Ref. \onlinecite{Ady_FMF},
the degeneracy of our bound states is not protected by a fundamental system property\cite{FK_theorem} and is susceptible to a specific kind of disorder.

The paper is organized as follows. In Sec. \ref{model} we introduce the model system; in Sec. \ref{majorana} we consider the non-inetracting case and find Kramers pairs of Majorana fermions, first for wires with SOI with opposite
 signs and then for wires with equal signs. 
 In Sec. \ref{Kramers_parafermion} we consider the case with interactions, and using a bosonization approach we derive the parafermion bound states. Finally, we give some conclusions in Sec. \ref{conclusions}.

\section{Model} 
\label{model}
\begin{figure}[!b]
\includegraphics[width=0.65\linewidth]{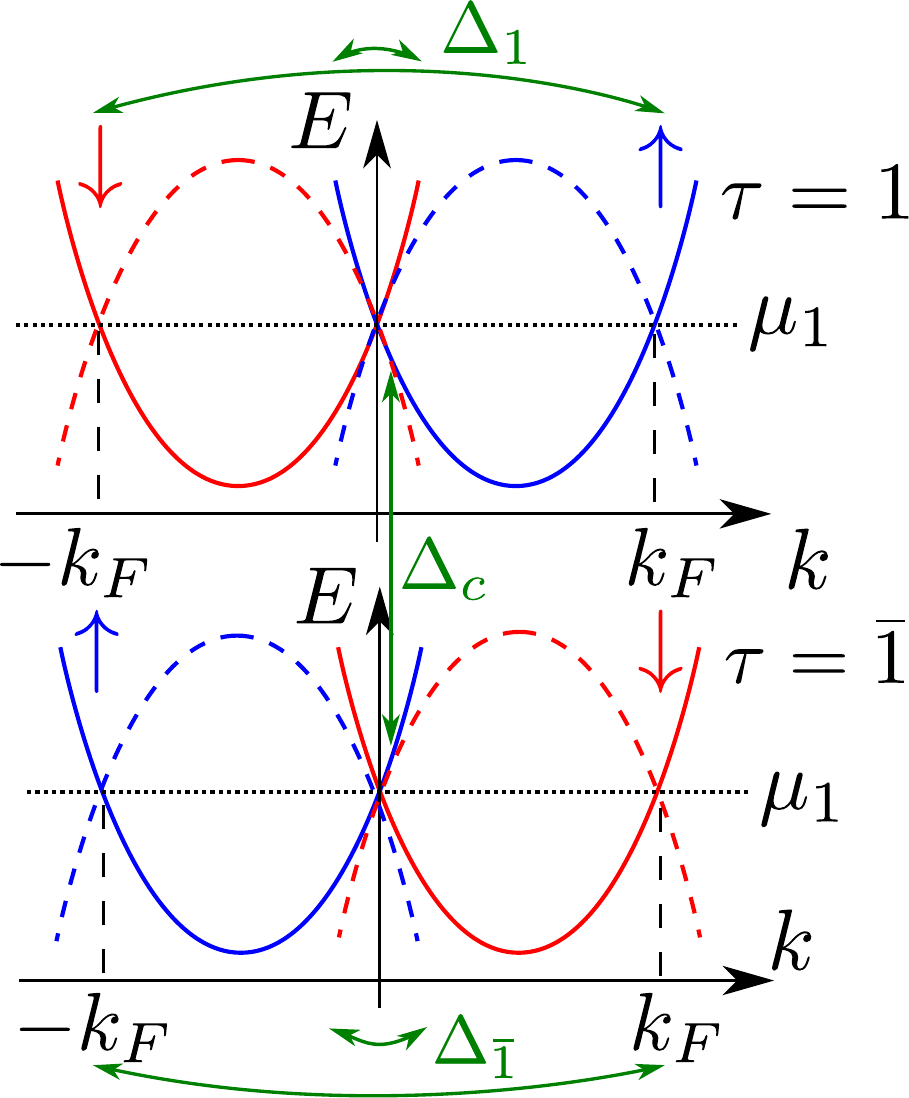}
\caption{The spectrum of two QWs with positive (negative) Rashba SOI  for $\tau=1$ ($\tau=\bar 1$). The solid (dashed) lines correspond to electrons (holes). The chemical potential $\mu$ is tuned to the crossing point between spin up (blue) and spin down (red).
The superconductivity couples states with opposite momenta and opposite spins belonging to the same $\tau$-wire ($\Delta_{\tau}$) and belonging to different wires ($\Delta_{c}$). The spectrum is gapless at $k=0$ for $\Delta_c^2=\Delta_1\Delta_{\bar 1}$, marking the topological phase transition that separates the topological phase with two localized midgap bound states at each wire end from the trivial phase without them.}
\label{spectrum}
\end{figure}

We consider a system consisting of two Rashba QWs brought into the proximity to an $s$-wave superconductor, see Fig.~\ref{bundle}. The upper (lower) QW is labeled by the index $\tau=1$ ($\tau=\bar1$) and is aligned in the $x$ direction. The kinetic part of the Hamiltonian is given by  
\begin{align}
H_0 =\sum_{\tau,\sigma}\int dx\ \Psi_{\tau\sigma}^\dagger(x) \left[ \frac{-\hbar^2 \partial_x^2}{2m}  - \mu_\tau \right]\Psi_{\tau\sigma}(x),
\end{align}
where $\Psi_{\tau\sigma}(x)^\dagger$ [$\Psi_{\tau\sigma}(x)$] is the creation (annihilation) operator  of an electron of mass $m$ at  position $x$ of the $\tau$-wire with  spin $\sigma/2 =\pm 1/2$ along the $z$-axis, and $\mu_\tau$ is the chemical potential.
The Rashba SOI field $\boldsymbol{\alpha}_{R\tau}$, that characterizes the strength and the direction of the spin polarization caused by SOI, points in the $z$ direction in each of the two QW, so the Rashba SOI term is written as
\begin{align}
H_{so}= - i \sum_{\tau,\sigma,\sigma'} \alpha_{R\tau} \int dx\ \tau \Psi_{\tau\sigma}^\dagger (\sigma_3)_{\sigma\sigma'} \partial_x\Psi_{\tau\sigma'}.
\end{align}
Here, the Pauli matrices $\sigma_{1,2,3}$ act on the spin of the electron. We note that the spin projection on the $z$ direction is a good quantum number ($\sigma$), and the 
dispersion relation 
for the spin component $\sigma$ at the $\tau$-wire is given by 
$E_{\tau\sigma} = \hbar^2 (k-\tau\sigma k_{so,\tau})^2/2m$,
where the chemical potential  $\mu$ is tuned to the crossing point between two spin-polarized bands at $k=0$, {\it i.e.} $\mu_1=E_{so}$, see Fig. \ref{spectrum}. Here, $E_{so,\tau}=\hbar^2k_{so,\tau}^2/2m$ is the SOI energy, and $k_{so,\tau}=m\alpha_{R\tau}/\hbar^2$ is the SOI wavevector.

In addition, the intrawire superconductivity of  strength $\Delta_\tau$ is proximity induced in each of the QWs by the tunneling of Cooper pairs as a whole from the superconductor to the $\tau$-wire. The corresponding pairing term  is given by
\begin{align}
H_{s}= \sum_{\tau,\sigma,\sigma'} \int dx\ \frac{\Delta_\tau}{2} [\Psi_{\tau\sigma}(i\sigma_2)_{\sigma\sigma'}\Psi_{\tau\sigma'}+H.c.].
\label{simplepairing1}
\end{align}
If the distance between two QWs is shorter than the superconductor coherence length then crossed Andreev reflection is possible \cite{Feinberg_2000}
where the electrons from the same Cooper pair tunnel into two different QWs, resulting in the interwire proximity induced superconductivity. \cite{Sasha,Two_LL_SC_Patrik,Andreev_Bena,Andreev_Yaroslav}  The corresponding pairing term is given by
\begin{align}
H_{c}=\sum_{\tau,\sigma,\sigma'} \int dx\ \frac{\Delta_{c}}{2}  [\Psi_{\tau\sigma}(i\sigma_2)_{\sigma\sigma'}\Psi_{\bar \tau\sigma'}+H.c.],
\end{align}
where $\Delta_{c}$ is the strength of the induced interwire superconductivity. 
Such a process is useful in Cooper pair splitters where 
crossed Andreev reflection dominates,\cite{Recher_Sukhorukov_Loss,Schonenberger,Heilblum}
so $\Delta_{c}>\Delta_{\tau}$.

Finally, we note that $H_{c}$ becomes equivalent to  FFLO pairing if one gauges away the SOI in the wires.
It is known that in one-dimensional wires the Rashba SOI can be gauged away by a spin-dependent gauge transformation. \cite{Braunecker} In our case, we gauge away the Rashba SOI simultaneously in both wires by the following transformation
\begin{equation}
\Psi'_{\tau \sigma}=e^{i\tau \sigma k_{so,\tau}x} \Psi_{\tau \sigma},
\end{equation}
which is also wire-dependent ($\tau$) as a consequence of opposite Rashba SOI.
As a result, the crossed Andreev term $H_{c}$ becomes in this new gauge
\begin{align}
&H'_{c}=\frac{1 }{2}\sum_{\tau,\sigma,\sigma'} \int dx \nonumber\\
& \Big[\Delta_{c} e^{-i\tau \sigma (k_{so,1}-k_{so,\bar 1})x} \Psi'_{\tau\sigma}(i\sigma_2)_{\sigma\sigma'}\Psi'_{\bar \tau\sigma'}+H.c.\Big ],
\end{align}
whereas $H_s$ remains unchanged. Thus, the crossed Andreev superconductivity has a non-uniform pairing term, $\Delta_{c} e^{-i\tau \sigma (k_{so,1}-k_{so,\bar 1}})x $, which manifestly breaks the translation invariance if $k_{so,1}\neq k_{so,\bar 1}$. This term is related to the Fulde-Ferrel-Larkin-Ovchinnikov (FFLO) state,\cite{FF,LO_1,LO_2} where the Cooper pair has finite total momentum. Therefore, all results derived in the main part for two wires with opposite Rashba SOI are also valid for a system consisting of two wires without SOI but coupled to  an FFLO-type superconductor instead of an ordinary $s$-wave superconductor.

The spatial dependence makes it explicit that there can be ground states in the system with broken symmetries (such as a charge density wave state), and thus states of different symmetries separated by domain walls that host bound states.
We note that this situation is analogous to  Ref. \onlinecite{Ady_FMF}, which finds parafermions in a one-dimensional Rashba wire coupled to a superconductor and in the presence of magnetic fields. There, it has been pointed out  \cite{Ady_FMF} that the resulting gapped state is not within
the list of possible gapped one-dimensional phases classified in Ref. \onlinecite{FK_theorem}.  As a consequence,  disorder or deviations from the mean-field description of superconductivity  can lift, in principle, the bound state degeneracy.  \cite{Ady_FMF}

\section{Kramers pairs of Majorana fermions} 
\label{majorana}

\subsection{SOIs of opposite sign}
\label{opposite_SOI}

 In this subsection we focus on the case where the Rashba SOIs are of opposite sign in the two QWs, $\alpha_{R1}\alpha_{R\bar1}<0$. In addition, the chemical potential is tuned to the SOI energy in both QWs, $\mu_\tau = E_{so,\tau}$. To simplify analytical calculations, we assume in what follows that $\alpha_{R1}=-\alpha_{R\bar1}=\alpha_R$. We note that the choice of exactly opposite SOIs, such that the Fermi velocities $\upsilon_F$ are the same in the two QWs,  is convenient but not necessary. All that is needed is  to tune the individual Fermi wavevectors $k_{F\tau}$ (via chemical potentials) to the individual $k_{so,\tau}$ values (or fractions thereof) in each wire. 

The proximity-induced superconductivity leads to gaps in the spectrum. Thus, the question arises if there are zero-energy bound states  localized at the ends of the wires. To find an answer, we proceed by linearizing the spectrum around the Fermi points $k=0$ and $k=\pm k_F\equiv \pm 2k_{so}$ (see Fig.~\ref{spectrum}),
\begin{align}
&\Psi_{11}=R_{11}e^{i k_F  x}+L_{11},\\
&\Psi_{1\bar 1}=L_{1\bar1}e^{-ik_Fx} + R_{1\bar1},\\
&\Psi_{\bar 11}=L_{\bar 11}e^{-ik_Fx} +R_{\bar 11},\\
&\Psi_{\bar1\bar 1}=R_{\bar 1\bar1} e^{ik_Fx}+L_{\bar 1\bar1},
\end{align}
where $R_{\tau\sigma}(x)$ [$L_{\tau\sigma}(x)$] are slowly varying right (left) mover fields of the electron with the spin $\sigma/2$ at the $\tau$-wire.\cite{Klinovaja2012,Rotating_field,Klinovaja_Loss_Ladder} Thus, $H_0+H_{so}$ reduces to
\begin{align}
&H_{kin}=i\hbar \upsilon_F \sum_{\tau,\sigma}\int dx [L_{\tau\sigma}^\dagger\partial_xL_{\tau\sigma} - R_{\tau\sigma}^\dagger\partial_xR_{\tau\sigma}],
\end{align}
and the superconductivity part to
\begin{align}
&H_{s}= \sum_{\tau} \int dx\  \frac{\Delta_{\tau}}{2} (R_{\tau1}^\dagger L_{\tau\bar1}^\dagger
- L_{\tau\bar1}^\dagger R_{\tau1}^\dagger 
\label{simplepairing2}
\nonumber\\
&\hspace{90pt}+L_{\tau1}^\dagger R_{\tau\bar1}^\dagger- R_{\tau\bar1}^\dagger L_{\tau1}^\dagger + H.c.
),\\
&H_{c}=\frac{\Delta_{c}}{2} \int dx\ ( L_{11}^\dagger L_{\bar 1\bar1}^\dagger - L_{\bar 1\bar1}^\dagger L_{11}^\dagger \nonumber\\
& \hspace{90pt} +R_{\bar 11}^\dagger R_{ 1\bar1}^\dagger -  R_{ 1\bar1}^\dagger R_{\bar 11}^\dagger + H.c.
).
\end{align}
Here, $\upsilon_F = \hbar k_F/m$ is the Fermi velocity. We note that the interwire superconductivity $\Delta_{c}$ couples only states with momenta close to zero,  see Fig.~\ref{spectrum}.

Combining together $H_{kin}$, $H_{s}$, and $H_{c}$, we arrive at the following Hamiltonian density $\mathcal{H}$, $H=(1/2) \int dx\ \hat\Psi ^\dagger(x) \mathcal{H} \hat\Psi (x)$,
\begin{align}
&\mathcal{H} =\hbar\upsilon_F \hat k \rho_3 +\Delta_{c} (\tau_1\eta_2\sigma_2+\tau_2\eta_2\sigma_1\rho_3)/2\nonumber\\
&\hspace{17pt}+\Delta_1 (1+\tau_3) \eta_2\sigma_2\rho_1/2 
+\Delta_{\bar 1} (1-\tau_3) \eta_2\sigma_2\rho_1/2,
\label{MF_hamiltonian_1}
\end{align}
where the basis is chosen to be $\hat\Psi$=($R_{11}$, $L_{11}$, $R_{1\bar1}$, $L_{1\bar1}$, $R^\dagger_{11}$, $L^\dagger_{11}$, $R^\dagger_{1\bar1}$, $L^\dagger_{1\bar1}$, $R_{\bar 1\bar1}$, $L_{\bar 1 1}$, $R_{\bar1\bar1}$, $L_{\bar1\bar1}$, $R^\dagger_{\bar 1\bar1}$, $L^\dagger_{\bar 1 1}$, $R^\dagger_{\bar1 \bar1}$, $L^\dagger_{\bar 1 \bar1})$. The Pauli matrices $\tau_{1,2,3}$ ($\sigma_{1,2,3}$) act in the QW (spin) space. The Pauli matrices $\eta_{1,2,3}$ ($\rho_{1,2,3}$) act in the electron-hole (right-left mover) subspace. The time-reversal operator $U_T= \sigma_2 \rho_1$ satisfies $U_T^\dagger {\mathcal H}^*(-k) U_T = {\mathcal H}(k)$. The particle-hole symmetry operator $U_P=\eta_1$ satisfies  $U_P^\dagger {\mathcal H}^*(-k) U_P = - {\mathcal H}(k)$. As a result, the system under consideration belongs to  topological symmetry class $\rm DIII$. \cite{Ryu}

The spectrum of the system is given by
\begin{align}
&E_{\tau,\pm}^2=(\hbar\upsilon_F k)^2+\Delta_\tau^2,\\
&E_{ 2,\pm,\pm}^2=\frac{1}{2}\Big(2(\hbar\upsilon_F k)^2+\Delta_1^2+\Delta_{\bar 1}^2+2\Delta_{c}^2\\
&\pm\sqrt{(\Delta_1^2-\Delta_{\bar 1}^2)^2+4\Delta_{c}^2[4(\hbar\upsilon_F k)^2+(\Delta_1+\Delta_{\bar 1})^2]}\Big)\nonumber,
\end{align}
where each level is twofold degenerate due to the time-reversal invariance of the system.
The system is gapless at $k=0$ if $\Delta_{c}^2=\Delta_{ 1}\Delta_{\bar 1}$ and at $k=\pm2\sqrt{\Delta_{c}^2 - \Delta_{ 1}^2}/\hbar\upsilon_F$  if $\Delta_{ 1}=\Delta_{\bar 1}<\Delta_{c}$.
In the latter case, the gap closes twice since the levels are twofold degenerate.
Although this does not change the number of  bound states, the supports of the corresponding wavefunctions are different.

Generally, if $\Delta_{c}^2>\Delta_{ 1}\Delta_{\bar 1}$ and $\Delta_{ 1}\neq\Delta_{\bar 1}$, there are  two zero-energy bound states localized at the left end and two at the right end of the system. These two states are Kramers partners protected by the time-reversal symmetry. Below we provide the wavefunction  $\Phi_{\rm MF1}(x)$ of one of these left-localized states written in the basis ($\Psi_{11}$, $\Psi_{1\bar{1}}$,$\Psi^\dagger_{11}$, $\Psi^\dagger_{1\bar1}$, $\Psi_{\bar 11}$, $\Psi_{\bar 1,\bar1}$, $\Psi^\dagger_{\bar 11}$,  $\Psi^\dagger_{\bar 1\bar1}$). Applying the time-reversal symmetry operator $T$, we find the wavefunction of its Kramers partner $\Phi_{\rm MF\bar 1}(x) = T\Phi_{\rm MF1}(x)$. The general form of the Majorana fermion wavefunction is then given by
\begin{align}
\Phi_{\rm MF1}(x)=\begin{pmatrix}
f_1(x)\\g_1(x)\\f_1^*(x)\\g_1^*(x)\\f_{\bar1}(x)\\g_{\bar1}(x)\\f_{\bar1}^*(x)\\g_{\bar1}^*(x)
\end{pmatrix}, \ \ \Phi_{\rm MF\bar1}(x)=\begin{pmatrix}
g_1^*(x)\\- f_1^*(x)\\g_1(x)\\-f_1(x)\\g_{\bar 1}^*(x)\\- f_{\bar 1}^*(x)\\g_{\bar 1}(x)\\-f_{\bar 1}(x),
\end{pmatrix}, \label{MF_general}
\end{align}
which follows from the requirement that the Majorana operators [belonging to zero-energy eigenstates of Eq. (\ref{MF_hamiltonian_1})] be self-adjoint: $\hat \Psi_{\rm MF 1}(x)=\hat\Psi^\dagger_{\rm MF 1}(x)$. From now on, without  loss of generality, we assume that $\Delta_{ 1}>\Delta_{\bar 1}$.

Next, we solve the eigenvalue equation for the Hamiltonian density given in Eq. (\ref{MF_hamiltonian_1}) for zero eigenenergy
 explicitly (following Ref. \onlinecite{Klinovaja2012}).
 If $\Delta_1+\Delta_{\bar 1}>2\Delta_{c}$,  the components of the corresponding wavefunctions are found to be given by 
\begin{align}
&f_1 (x)= -i g^*_1(x)= (e^{-x/\xi_2}-e^{-x/\xi_{\bar 2} })\\
&\hspace{15pt}\times\Delta_{c}\left(\Delta_1+\Delta_{\bar 1}+\sqrt{(\Delta_1+\Delta_{\bar 1})^2-4\Delta_{c}^2}\right),\nonumber \\
&f_{\bar{1}} (x)= -i g^*_{\bar{1}}(x)= -2 \Delta_{c}^2 e^{-x/\xi_2}\\
&-e^{-x/\xi_{\bar1}+ik_Fx} \sqrt{(\Delta_1+\Delta_{\bar 1})^2-4\Delta_{c}^2}\nonumber\\
& \hspace{40pt}\times  \left(\Delta_1+\Delta_{\bar 1}+ \sqrt{(\Delta_1+\Delta_{\bar 1})^2-4\Delta_{c}^2}\right) \nonumber\\
&+\frac{1}{2}e^{-x/\xi_{\bar 2}}\left(\Delta_1+\Delta_{\bar 1}+\sqrt{(\Delta_1+\Delta_{\bar 1})^2-4\Delta_{c}^2}\right)^2\nonumber,
\end{align}
where the localization lengths are given by
\begin{align}
&\xi_{\pm 1}=\hbar\upsilon_F/\Delta_{\pm 1},\\
& \xi_{\pm 2}=2\hbar\upsilon_F /\left(\Delta_1-\Delta_{\bar 1}\pm\sqrt{(\Delta_1+\Delta_{\bar 1})^2-4\Delta_{c}^2}\right).\nonumber
\end{align}
If $\Delta_1+\Delta_{\bar 1}<2\Delta_{c}$, the wavefunction components are given by
\begin{align}
&f_1 (x)= i g^*_1(x)=  -e^{-x/\xi_3} \sin (k_1 x)  \\
&\hspace{15pt}\times 2 \Delta_{c}\left(\Delta_1+\Delta_{\bar 1}+i\sqrt{4\Delta_{c}^2-(\Delta_1+\Delta_{\bar 1})^2}\right),\nonumber \\
&f_{\bar{1}} (x)= i g^*_{\bar{1}}(x)\\
&= e^{-x/\xi_{\bar1}+ik_Fx}\sqrt{4\Delta_{c}^2-(\Delta_1+\Delta_{\bar 1})^2}\nonumber\\
&\hspace{40pt}\times  \left(\Delta_1+\Delta_{\bar 1}+i \sqrt{4\Delta_{c}^2-(\Delta_1+\Delta_{\bar 1})^2}\right) \nonumber\\
& - e^{-x/\xi_3} \left(\Delta_1+\Delta_{\bar 1}
+ i \sqrt{4\Delta_{c}^2-(\Delta_1+\Delta_{\bar 1})^2}\right)\nonumber\\
&\times\Big[ \cos(k_1x) \sqrt{4\Delta_{c}^2-(\Delta_1+\Delta_{\bar 1})^2} +\sin(k_1x)(\Delta_1+\Delta_{\bar 1})  \Big],\nonumber
\end{align}
where the localization length $\xi_{\pm 3}$ and the wavevector $k_{1}$ are given by
\begin{align}
& \xi_{3}=2\hbar\upsilon_F /\left(\Delta_1-\Delta_{\bar 1}\right),\\
&k_{1}=\pm\sqrt{4\Delta_{c}^2-(\Delta_1+\Delta_{\bar 1})^2}/2\hbar\upsilon_F .\nonumber
\end{align}

The case of $\Delta_1+\Delta_{\bar 1}=2\Delta_{c}$ should be treated separately leading to
\begin{align}
&f_1(x)=-ig_1^*(x)=-i x e^{-x/\xi_3} (\Delta_1 + \Delta_{\bar 1})/\hbar\upsilon_F ,\\
&f_{\bar 1}(x)=-ig_{\bar 1}^*(x)=-i\Big(2e^{-x/\xi_{\bar 1}+ik_Fx}\nonumber\\
&\hspace{60pt}-e^{-x/\xi_{3}}[2+x(\Delta_1 + \Delta_{\bar 1})/\hbar\upsilon_F]\Big).
\end{align}

\begin{figure}[!b]
\includegraphics[width=0.65\linewidth]{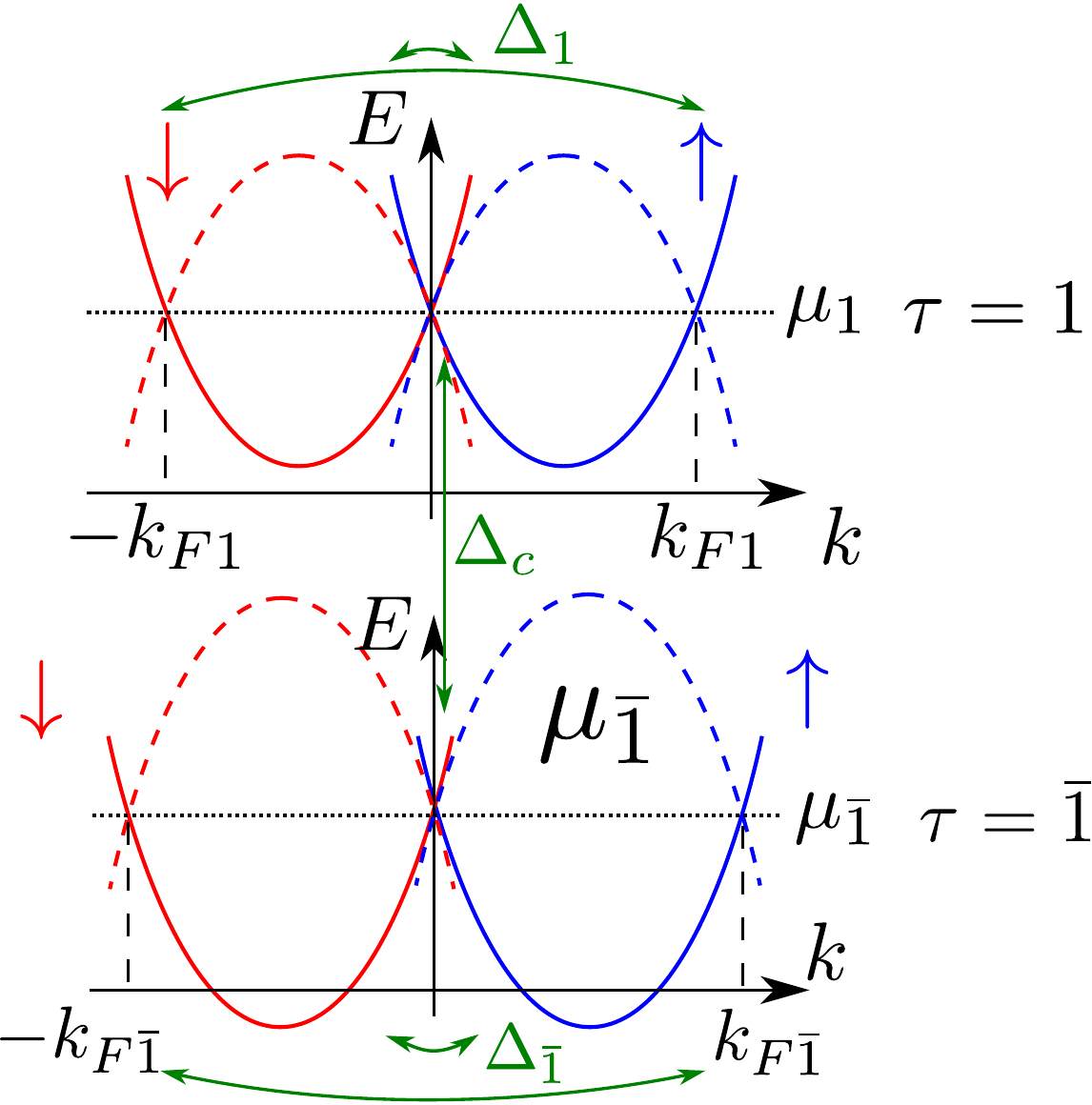}
\caption{The spectrum of two QWs with positive Rashba SOI in both QWs. The solid (dashed) lines correspond to electrons (holes). The chemical potential $\mu_\tau$ is tuned to the crossing point between spin up (blue) and spin down (red).
The superconductivity couples states with opposite momenta and opposite spins belonging to the same $\tau$-wire ($\Delta_{\tau}$) and belonging to different wires ($\Delta_{c}$). The spectrum is gapless at $k=0$ for $\Delta_c^2=\Delta_1\Delta_{\bar 1}$, marking the topological phase transition that separates the topological phase with two localized midgap
bound states at each wire end from the trivial phase without them.}
\label{same_SOI}
\end{figure}

 As a result, if $\Delta_{c}^2>\Delta_{ 1}\Delta_{\bar 1}$ and $\Delta_{ 1}\neq\Delta_{\bar 1}$, we find two zero-energy bound states at each system end, and we denote the corresponding  Majorana operators (say, at the left end) as  $\Psi_{\rm MF\tau}= \Psi_{\rm MF\tau}^\dagger$. These MFs are  Kramers partners of each other, so that their wavefunctions are related by $\Phi_{\rm MF\bar 1}(x) = T\Phi_{\rm MF1}(x)$.  Here, the time-reversal operator $T$ is given by $T=i\sigma_2 K$, where  $K \Phi(x) =\Phi^*(x)$.

\subsection{SOIs of equal sign}
\label{equal_SOI}

In this subsection, we consider the case where the two QWs have the same sign of Rashba SOI, $\alpha_{R1}\alpha_{R\bar1}>0$. However, in this case, in contrast to the previous section, we assume that $\alpha_{R1}>\alpha_{R\bar1}>0$. Otherwise, as mentioned above, the SOI can be gauged away completely without generating the position-dependent  crossed Andreev pairing. Again, MFs emerge as a result of a competition between two pairing terms, and, importantly, the crossed Andreev pairing is possible only at $k=0$ but not at finite momenta, where states with opposite spins do not have opposite momenta, see Fig. \ref{same_SOI}. 

In this subsection we use the same notation for Hamiltonian as in the previous one. We believe that this should not lead to any misinterpretation but could help to make connections between two setups. In addition, taking into account that calculations are very similar in the two case, we try to keep the discussion short and omit details.

Again, we linearize the spectrum around the Fermi points  $k=0$ and $k_{F\tau}= \pm 2k_{so,\tau}$,
\begin{align}
&\Psi_{11}=R_{11}e^{i k_{F1}  x}+L_{11},\\
&\Psi_{1\bar 1}=L_{1\bar1}e^{-ik_{F1}x} + R_{1\bar1},\\
&\Psi_{\bar 1 1}=L_{\bar 11} +R_{\bar 11} e^{ik_{F\bar 1}x},\\
&\Psi_{\bar1\bar 1}=R_{\bar 1\bar1} +L_{\bar 1\bar1} e^{-ik_{F\bar 1}x}.
\end{align}
where $R_{\tau\sigma}(x)$ [$L_{\tau\sigma}(x)$] are slowly varying right (left) mover fields of the electron with the spin $\sigma/2$ at the $\tau$-wire.\cite{Klinovaja2012,Rotating_field,Klinovaja_Loss_Ladder} Here, we again assume that the chemical potentials are tuned to the SO energy, $\mu_\tau = E_{so,\tau}$.

The kinetic part of the Hamiltonian $H_0+H_{so}$ reduces to
\begin{align}
&H_{kin}= \sum_{\tau,\sigma}\int dx\ i\hbar \upsilon_{F\tau} [L_{\tau\sigma}^\dagger\partial_xL_{\tau\sigma} - R_{\tau\sigma}^\dagger\partial_xR_{\tau\sigma}],
\end{align}
and the superconductivity part to
\begin{align}
&H_{s}= \sum_{\tau} \int dx\  \frac{\Delta_{\tau}}{2} (R_{\tau1}^\dagger L_{\tau\bar1}^\dagger
- L_{\tau\bar1}^\dagger R_{\tau1}^\dagger 
\label{simplepairing2}
\nonumber\\
&\hspace{90pt}+L_{\tau1}^\dagger R_{\tau\bar1}^\dagger- R_{\tau\bar1}^\dagger L_{\tau1}^\dagger + H.c.
),\\
&H_{c}=\frac{\Delta_{c}}{2} \int dx\ ( L_{\bar 11}^\dagger R_{ 1\bar1}^\dagger - R_{ 1\bar1}^\dagger L_{\bar 11}^\dagger \nonumber\\
& \hspace{90pt} +L_{11}^\dagger R_{ \bar1\bar1}^\dagger -  R_{\bar 1\bar1}^\dagger L_{ 11}^\dagger + H.c.
).
\end{align}
Here, $\upsilon_{F\tau} = \hbar k_{F\tau}/m$ is the Fermi velocity. Again, the interwire superconductivity $\Delta_{c}$ acts only at momenta close to zero,  see Fig.~\ref{same_SOI}.

The Hamiltonian density $\mathcal{H}$ in terms of Pauli matrices is given by
\begin{align}
&\mathcal{H} =\hbar\upsilon_{F1} \hat k  (1+\tau_3) \rho_3 /2 +\hbar\upsilon_{F\bar 1} \hat k (1-\tau_3) \rho_3/2 \nonumber\\
&\hspace{50pt}+\Delta_{c} \tau_1\eta_2 (\sigma_1 \rho_2-\sigma_2\rho_2)/2\nonumber\\
&\hspace{15pt}+\Delta_1 (1+\tau_3) \eta_2\sigma_2\rho_1/2 
+\Delta_{\bar 1} (1-\tau_3) \eta_2\sigma_2\rho_1/2,
\label{MF_hamiltonian_1}
\end{align}
where the basis is chosen to be $\hat\Psi$=($R_{11}$, $L_{11}$, $R_{1\bar1}$, $L_{1\bar1}$, $R^\dagger_{11}$, $L^\dagger_{11}$, $R^\dagger_{1\bar1}$, $L^\dagger_{1\bar1}$, $R_{\bar 1\bar1}$, $L_{\bar 1 1}$, $R_{\bar1\bar1}$, $L_{\bar1\bar1}$, $R^\dagger_{\bar 1\bar1}$, $L^\dagger_{\bar 1 1}$, $R^\dagger_{\bar1 \bar1}$, $L^\dagger_{\bar 1 \bar1})$. The energy spectrum is given by
\begin{widetext}
\begin{align}
&E_{\tau,\pm}^2=(\hbar \upsilon_{F\tau} k)^2 + \Delta_\tau^2,\\
&E_{2,\pm,\pm}^2 = \frac{1}{2} \Big( \Delta_1^2+ \Delta_2^2+2 \Delta_c^2 + \hbar^2 ( \upsilon_{F1}^2+\upsilon_{F2}^2) k^2\\
&\pm \sqrt{(\Delta_1^2- \Delta_2^2)^2 + 4 \Delta_c^2 (\Delta_1+\Delta_{\bar 1})^2+ \hbar^4 ( \upsilon_{F1}^2-\upsilon_{F2}^2)^2 k^4
+4 \Delta_c^2 \hbar^2  (\upsilon_{F1}-\upsilon_{F2})^2 k^2 + 2 \hbar^2  (\upsilon_{F1}^2-\upsilon_{F2}^2) (\Delta_1^2-\Delta_{\bar 1}^2)k^2 }\Big), \nonumber
\end{align}
\end{widetext}
where each level is twofold degenerate.  We note again that the spectrum is gapless at $k=0$ provided that $\Delta_c^2 = \Delta_1 \Delta_{\bar 1}$. If $\Delta_{c}^2>\Delta_{ 1}\Delta_{\bar 1}$, we find two zero-energy bound states at each system end. The corresponding MF wavefunctions are too involved to be displayed in a general case. However, in the special simplified case with $\Delta_1=\Delta_{\bar 1}$ and $\upsilon_{F1}=\upsilon_{F2}$, the MFs are defined  by Eq. (\ref{MF_general}) with
\begin{align}
&f_1(x)= i g_1^*(x)=(e^{-ik_{F1}x} e^{-x/\xi_1} - e^{-x/\xi_2}),\\
&f_{\bar 1}(x)= i g_{\bar 1}^*(x) =(e^{-x/\xi_2} - e^{-ik_{F\bar 1}x} e^{-x/\xi_1} )
\end{align}
The localization length are given by $\xi_1=\hbar\upsilon_F/\Delta_1$ and $\xi_2=\hbar\upsilon_F/(\Delta_c -\Delta_1)$.

\section{Kramers pairs of parafermions}  
\label{Kramers_parafermion}

Electron-electron interaction effects  become important if the chemical potential is tuned to be, for example,  at one third of the SOI energy, $\mu_{1/3,\tau}=E_{so,\tau}/9$, such that the Fermi wavevectors become $\pm k_{so,\tau}(1\pm1/3)$. 
In this case, the interwire pairing  is possible only if  backscattering terms of finite strength $g_B$ are taken into 
account to generate momentum-conserving terms. \cite{Ady_FMF,Giamarchi,kane_PRL,kane,QHE_Klinovaja_PRL,QHE_Klinovaja_PRB}
Below, we focus on the second case of Rashba SOI of the same sign in both QWs.

\begin{figure}[t!]
\includegraphics[width=\linewidth]{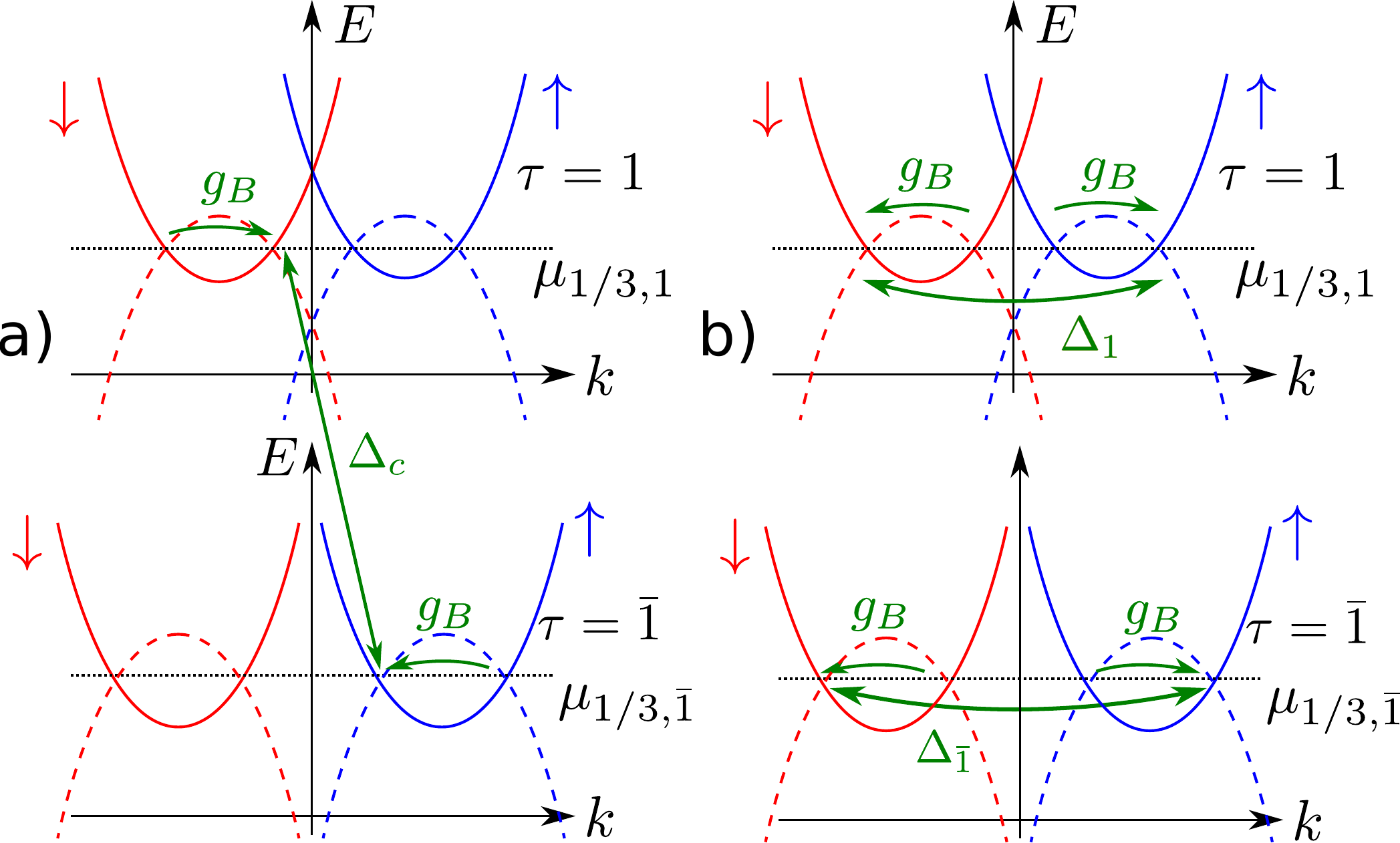}
\caption{The momentum-conserving scattering events corresponding to a) ${\mathcal H}_{c}^{ee}$ and b) ${\mathcal H}_{s,\tau}^{ee}$ for the chemical potential $\mu_{1/3,\tau}=E_{so,\tau}/9$  with associated Fermi wavevectors $\pm k_{so}(1\pm1/3)$.  See the caption of Fig. 2 for notations. }
\label{figure13}
\end{figure}

 In particular, the  interwire  superconductivity Hamiltonian density in Nambu space is given by
\begin{align}
&{\mathcal H}_{c}^{ee}=g_{c}\Big[ L_{\bar 11}^\dagger R_{ 1\bar1}^\dagger ( L_{\bar 11}^\dagger R_{ 11}) (L_{ 1\bar1}R_{ 1\bar1}^\dagger)  - R_{ 1\bar1}^\dagger L_{\bar 11}^\dagger  \nonumber\\
&\hspace{15pt}\times( R_{ 1\bar1}^\dagger L_{1\bar1}  ) (R_{ \bar 11} L_{\bar 11}^\dagger) + L_{11}^\dagger R_{ \bar1\bar1}^\dagger (L_{11}^\dagger R_{ 11})(L_{ \bar1\bar1} R_{ \bar1\bar1}^\dagger) \nonumber\\
&-  R_{\bar 1\bar1}^\dagger L_{ 11}^\dagger (R_{\bar 1\bar1}^\dagger L_{\bar 1\bar1})(R_{11} L_{ 11}^\dagger)+H.c.\Big],
\end{align}
where the coupling strength is given by $g_{c}\propto\Delta_{c} g_B^2 $. The structure of ${\mathcal H}_{c}^{ee}$ can be understood as follows. If a Cooper pair splits and each partner tunnels into a different QW ({\it i.e.} $L_{\bar 11}^\dagger R_{\bar 1\bar1}^\dagger$), both electrons go to the same momentum $k_F$, as a result, the finite momentum of such a Cooper pair should be compensated by two back-scattering events taking place inside each of the QWs ({\it i.e.} $ L_{\bar 11}^\dagger R_{ 11}$ and $L_{ \bar1\bar1}R_{\bar 1\bar1}^\dagger$). 

Next, we note that ${\mathcal H}_{c}^{ee}$ and ${\mathcal H}_{s}$ [defined by Eq.~(\ref{simplepairing2})] do not commute, so these two terms cannot be ordered simultaneously in the bosonized represenation (see below). Thus, only of these term can be dominant and result in the energy gap. In what follows, we assume that our setup is in the regime where ${\mathcal H}_{c}^{ee}$ dominates over ${\mathcal H}_{s}$. This corresponds to two possible cases: the scaling dimension $K_c$ of ${\mathcal H}_{c}^{ee}$ is the lowest one or the bare coupling constant $g_{c}$ is of order one. The scaling dimension $K_c = [K_\alpha^{-1}+K_\delta^{-1} + 9(K_\beta+K_\gamma) ]/4$ can be found in a usual way in the basis of conjugated bosonic fields $\phi_{\alpha,\beta,\gamma,\delta}$ and $\theta_{\alpha,\beta,\gamma,\delta}$:  $\chi_{r\tau\sigma}= [r \phi_\alpha + \theta_\alpha + \tau (r \phi_\beta +  \theta_\beta)+\sigma (r \phi_\gamma +  \theta_\gamma + \tau (r \phi_\delta +  \theta_\delta)]/2$. Here, the bosonic field $\chi_{1\
tau\sigma}$ ($\chi_{\bar1\tau\sigma}$) corresponds to the fermion operator $R_{\tau\sigma}$ ($L_{\tau\sigma}$). The scaling dimension of ${\mathcal H}_{s}$ is given in the same basis by $K_s=[K_\alpha^{-1}+K_\beta^{-1}+K_\gamma+ K_\delta ]/4$. Comparing $K_s$ and $K_c$, we see that in the regime of strong electon-electron interaction when the Luttinger parameters are substantially smaller than one, the crossed Andreev pairing is dominant, $K_s<K_c$.

The intrawire pairing term ${\mathcal H}_{s}^{ee} =\sum_\tau {\mathcal H}_{s,\tau}^{ee}$ that commutes with ${\mathcal H}_{c}^{ee}$ is given by
\begin{align}
&{\mathcal H}_{s,\tau }^{ee}=
g_\tau \Big[R_{\tau 1}^\dagger L_{\tau  \bar 1}^\dagger (R_{\tau 1}^\dagger L_{\tau 1})(R_{\tau \bar 1} L_{\tau \bar 1}^\dagger )
\nonumber\\
&\hspace{55pt} -L_{\tau \bar1}^\dagger R_{\tau 1}^\dagger(L_{\tau 1} R_{\tau 1}^\dagger )(L_{\tau \bar1}^\dagger R_{\tau \bar1})
+H.c. \Big],
\end{align}
where 
$g_\tau\propto\Delta_\tau g_B^2$.

Next, we perform a bosonization of the fermions~\cite{Giamarchi} in  Nambu space. For this we represent electron (hole) operators as $R_{\tau \sigma}=e^{i\phi_{1\tau\sigma}}$ and $L_{\tau \sigma}=e^{i\phi_{\bar1\tau\sigma}}$  ($R^\dagger_{\tau \sigma}=e^{i\tilde\phi_{1\tau\sigma}}$ and $L^\dagger_{\tau \sigma}=e^{i\tilde\phi_{\bar1\tau\sigma}}$)
in terms of chiral fields $\phi_{r\tau\sigma}$ and $\tilde\phi_{r\tau\sigma}$, where  $r$ refers to the right/left movers, and $\tau$ ($\sigma$) labels the QW (spin). 
 We then get,
\begin{align}
{\mathcal H}_{c}^{ee}=&2g_{c}   \Big[\cos(2\phi_{\bar 1 \bar 1 1} -2\tilde \phi_{1 1 \bar 1} -\phi_{ 1 \bar 1 1} +\tilde \phi_{\bar 1  1 \bar 1} )\nonumber\\
&\hspace{15pt}-\cos(2\phi_{11 \bar 1}-2\tilde\phi_{\bar 1 \bar1  1}-\phi_{\bar 11 \bar 1}+\tilde\phi_{ 1 \bar1  1})\nonumber\\
&\hspace{15pt} \cos(2\phi_{\bar 1  1 1} -2\tilde \phi_{1 \bar 1 \bar 1} -\phi_{ 11 1} +\tilde \phi_{\bar 1 \bar 1 \bar 1} )\nonumber\\
&\hspace{15pt}-\cos(2\phi_{1 \bar 1 \bar 1}-2\tilde\phi_{\bar 1 1  1}-\phi_{\bar 1 \bar 1 \bar 1}+\tilde\phi_{  \bar 1 1  1})
\Big],\label{H_cN}\\
{\mathcal H}_{s,\tau}^{ee}=&2g_{\tau} [\cos(2\phi_{1 \tau 1}-2\tilde {\phi}_{\bar1  \tau \bar 1}-\phi_{\bar 1 \tau1}+\tilde {\phi}_{1 \tau \bar 1})\nonumber\\
&\hspace{15pt}-\cos(2\phi_{\bar 1 \tau \bar 1}-2\tilde {\phi}_{1 \tau 1}-\phi_{ 1 \tau \bar 1}+\tilde {\phi}_{\bar 1 \tau 1})]. \label{H_sN}
\end{align}

Next, we separate the total Hamiltonian into two uncoupled commuting parts, $H+ {\bar H}$, where
$H$ ($\bar H$) operates in the space spanned by $(\phi_{r\tau 1},\tilde\phi_{ r\tau\bar 1})$ [$(\phi_{r\tau \bar1},\tilde\phi_{ r\tau 1})$]. 
Thus, $H$ and $\bar H$ operate in time-reversal conjugated spaces, which we can treat as two independent subsystems. Thus, we will focus only on $H$, knowing that the solution for $\bar H$ can be obtained by direct analogy or via the requirement of time-reversal symmetry. To simplify calculations, we introduce new notations $\eta_{r\tau\sigma}=2\phi_{r\tau\sigma}-\phi_{\bar r\tau\sigma}$ and $\tilde\eta_{r\tau\sigma}=2\tilde\phi_{r\tau\sigma}-\tilde\phi_{\bar r\tau\sigma}$. This results in
\begin{align}
&{\mathcal H}^{ee}=2g_1  \cos(\eta_{111}-\tilde {\eta}_{\bar1 1 \bar 1})+2g_{\bar 1} \cos(\eta_{1 \bar 1 1}-\tilde\eta_{\bar 1\bar 1\bar1})\nonumber\\
&\hspace{10pt}+2g_{c} \cos(\eta_{\bar 1 \bar 1 1}-\tilde\eta_{1  1 \bar 1})+2g_{c}   \cos(\eta_{ \bar 1  1  1}-\tilde\eta_{ 1  \bar 1  \bar 1}).
\end{align}
Searching for bound states, we  impose vanishing boundary conditions at $x=0,\ell$, which couples right and left movers,
$\eta_{1\tau\sigma} (x=0,\ell) = \eta_{\bar 1\tau \sigma }(x=0,\ell) +\pi$.
Next, we unfold the QWs~\cite{LL_Eggert,LL_Gogolin,Giamarchi,PF_Clarke,Ady_FMF,FF_suhas,MF_ee_Suhas} by formally extending them from $-\ell$ to $\ell$ by defining new chiral fields such that the boundary conditions are satisfied automatically,
\begin{align}
&
\xi_{\tau}(x)=\begin{cases} \eta_{1 \tau 1}(x), & x>0\\
\eta_{\bar 1  \tau 1}(-x)+\pi, & x<0\, ,
\label{chiralfield}
\end{cases}
\end{align}
and analogously we define $\tilde\xi_\tau$ with $\tilde\eta$'s.
Next, we transform the chiral fields to conjugate fields $\phi, \theta$, via
$\xi_{\tau} = ( \phi_1+  \theta_1 +3 \tau \phi_2  +3 \tau \theta_2)/2 $
and $\tilde\xi_{r} = (-\phi_1+ \theta_1 - 3 \tau  \phi_2  + 3 \tau \theta_2)/2$.
Finally, we arrive at
\begin{align}
&{\mathcal H}^{ee}=\begin{cases}
2\sum_{\tau}g_\tau \cos(\phi_1+ 3\tau\phi_2), & \hspace{0pt} x>0\\
4g_{c} \cos (\phi_1) \cos (3\theta_2),
&\hspace{0pt} x<0\, .
\end{cases}\label{Nambu_ee}
\end{align}
Working in the limit of strong electron-electron interactions, we assume that $g_{\tau}$  and $g_{c}$ are large enough, so that the interaction terms are dominant, resulting in the pinning of the fields to constant values such that the total energy is minimized. \cite{PF_Linder,PF_Clarke,Ady_FMF,PF_Mong,vaezi_2,PFs_Loss,bilayer_PFs} Thus, we conclude that the field $\phi_1 =\pi M$ is pinned uniformly to minimize the kinetic energy. In addition, the two non-commuting conjugated fields $\theta_2$ and $\phi_2$ are pinned in two neighbouring regions separated by an infinitesimal interval,
\begin{align}
&\theta_2 = \pi(1+M+2m)/3,\ \ \ x<0,\\
&\phi_2 = \pi(1+M+2n)/3,\ \ x>0,
\end{align}
where $M$, $n$, and $m$ are integer-valued operators. We note that the only non-zero commutator is $[m,n]=3i/4\pi$, which follows directly from 
$[\phi_{2}(x), \theta_{2}(x')]=-(i\pi/3){\rm sgn}(x-x')$, which in turn follows from the standard commutation relation for the chiral fields $\xi_{\tau}$ and $\tilde \xi_{\tau}$
defined in Eq.~(\ref{chiralfield}).
Next, we  define two operators that commute with the Hamiltonian, so that they correspond to zero energy states,
\begin{align}
\alpha_{1}=e^{ i\frac{4\pi}{3} (m-n)},\ \alpha_{\bar 1}=e^{ i\frac{4\pi}{3} (m+n)}.
\end{align}
These operators act at the QW ends \cite{PF_Clarke} and are easily seen to satisfy $\alpha_{1}^3=\alpha_{\bar 1}^3=1$, and  $\alpha_{1}\alpha_{\bar 1}=\alpha_{\bar 1} \alpha_{1}e^{-2i\pi /3}$.
Thus, they form  parafermions.  
We further note that the ground state of $H$ is threefold degenerate. Indeed, from $(\alpha^\dagger_{1} \alpha_{\bar 1})^3=1$ we see that $\alpha^\dagger_{1} \alpha_{\bar 1}$ has three distinct eigenvalues $e^{2i\pi q /3}$, where  
$q=0,\pm 1 \ ({\rm mod}\ 3) $.
The corresponding eigenstates are denoted by $\ket{q}$.
With an appropriate phase choice, \cite{PF_Clarke}
we find
$\alpha_{1} \ket{q} = \ket{q+1}$, 
so the ground state is threefold degenerate in the considered subspace.
Analogously, we obtain the Kramers partners from $\bar H$, $\bar\alpha_\tau=\alpha_\tau(m,n \rightarrow {\bar m}, {\bar n})$, where,  again, $\bar m$, and $\bar n$ are integer-valued operators, and  $\bar q=0,\pm 1 \ ({\rm mod}\ 3) $. Thus, the ground state of the entire system, $\ket{q}\otimes\ket{\bar q}$, consist of three Kramers pairs of parafermions. However, as shown in Ref. \onlinecite{Ady_FMF}, the degeneracy could be lifted by  disorder.
As a result, the parafermion phase found here does not belong to the topological phases classified in \onlinecite{FK_theorem}.

We note that due to our basis choice the constructed states $\ket q,\ket {\bar q}$  are not particle-hole symmetric. However, one can easily find new particle-hole invariant states by combining two Kramers partners with appropriate phase.

\begin{figure}[h!]
\includegraphics[width=0.6\linewidth]{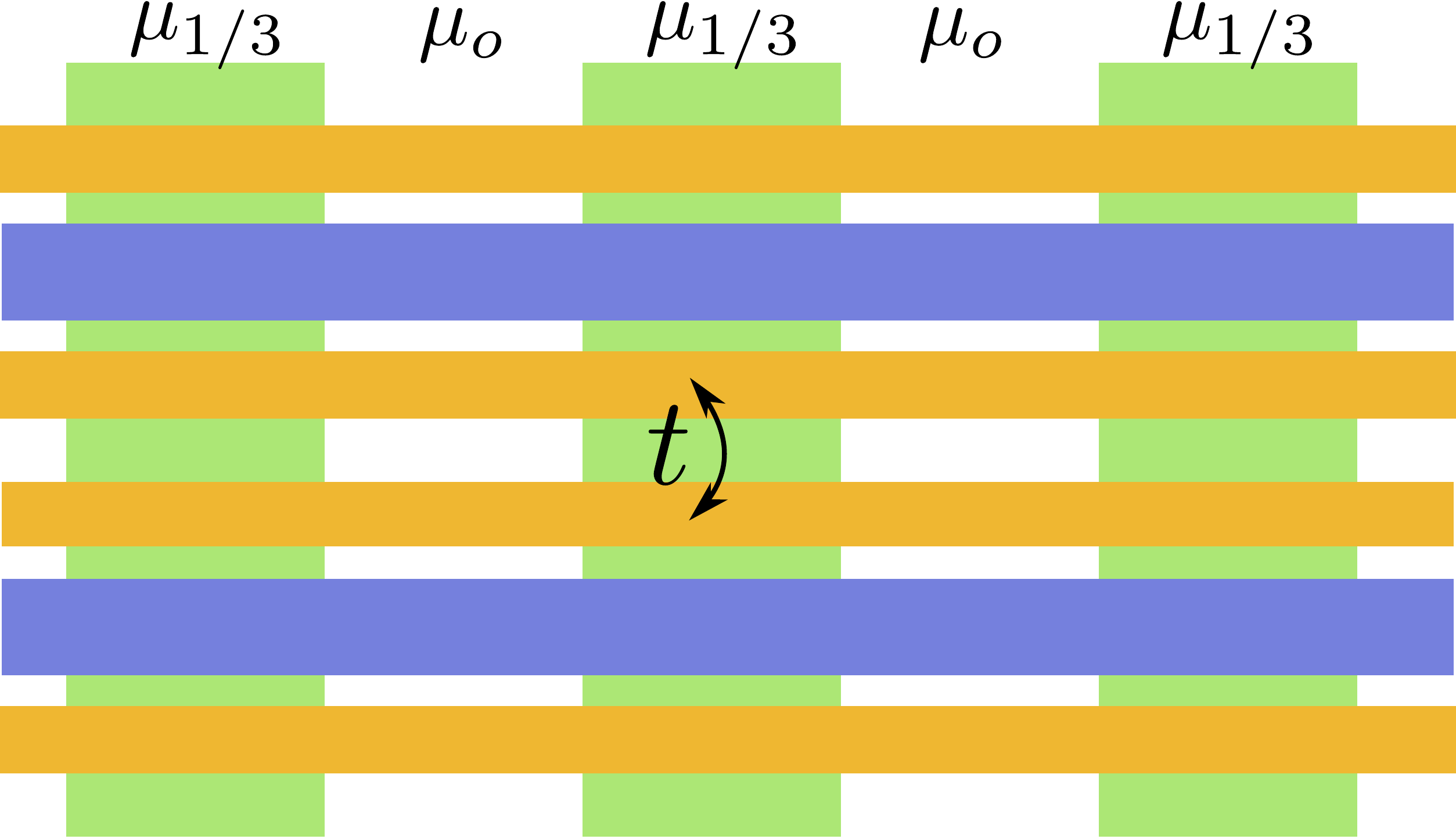}
\caption{Two-dimensional system of parafermions consisting of an array of coupled QWs with proximity induced interwire and intrawire superconductivity, see Fig. 1 in the main part.
The transition between the interwire-pairing-dominant phase ($\mu_{1/3}$) and the intrawire-pairing-dominant  phase ($\mu_o$) is controlled by electrical gates (green slabs).
Parafermions are formed initially at the boundaries between these two phases.  The tunneling $t$ between two neighbouring QWs not separated by a superconductor results in  deconfinement~\cite{PF_Mong} and in a sea of time-reversal invariant parafermions.}
\label{2D}
\end{figure}

So far  we have considered QWs of finite length which are entirely in the 
non-trivial phase supporting PFs localized at the wire ends. However, by local tuning of the chemical potential $\mu$, we can move parafermions inside the QWs, see Fig. \ref {2D}. As shown above, if $\mu=\mu_{1/3}$, the interwire superconductivity dominate.
 However, if $\mu=\mu_o$ is significantly detuned from  $\mu_{1/3}$, the interwire superconductivity $H_{c}^{ee}$ is suppressed. Thus, the intrawire superconductivity $H_{s}$ dominates, driving this part of the system into the trivial phase. As before, PFs are localized at the boundary between two phases. All this allows us to generate PF networks that can also extend to two-dimensional setups. \cite{Alicea_braiding}
Introducing coupling between parafermions
one generates a sea of PFs, which can potentially result in the Fibonacci phase
as argued in Ref.~\onlinecite{PF_Mong}. At the same time, the extension to a two-dimensional system can help to stabilize this phase and make it less susceptible to disorder. 

The presence of the PFs in the gap can be tested in setups similar to the ones developed for MFs. \cite{mourik_signatures_2012,deng_observation_2012,das_evidence_2012,Rokhinson,Goldhaber,marcus_MF} In particular,  one can detect PFs  by the zero bias peak in the conductance. 
The periodicity of the Josephson current as  function of the superconducting phase provides more information. As shown before, the period for ${\mathbb Z}_n$ PFs is $2\pi n$. \cite{PF_Clarke} For time-reversal invariant PFs, similar to  time-reversal invariant  MFs, \cite{TRI6} several periods can be observed with $2\pi n$ being the largest one, {\it i.e.,} $6\pi$ for the PFs considered in this work.

\section {Conclusions} 
\label{conclusions}

We showed that it is possible to construct Kramers pairs of PFs in a time-reversal invariant system. 
As an example of such a setup we considered Rashba QWs coupled to a superconductor. Given the rapid experimental progress with similar ultraclean systems designed for MFs,\cite{mourik_signatures_2012,deng_observation_2012,das_evidence_2012,marcus_MF} the proposed setup seems to be within experimental reach.
In addition, we mention that a similar scheme  works also for edge states of fractional topological insulators (or fractional quantum spin Hall effect system), where different topological regions can be induced by superconductivity and transverse hopping. 
We also envisage the extension of our system to a 2D network  that might result in a Fibonacci phase. \cite{PF_Mong}  The construction of quantum gates for time-reversal invariant parafermions is an interesting problem by itself, and could be addressed in further  work. We also leave for further work a study of the splitting potentially caused by disorder effects. 
However, we envisage that if disorder effects lift the degeneracy of the bound states, the resulting energy splitting of states can serve as a useful tool to experimentally access the level of the initial ground state degeneracy, such that we can distinguish directly one Kramers pair of MFs from three Kramers pairs of PFs.

\acknowledgments
We thank the UCSB KITP for hospitality.
This research is supported by the  Harvard Quantum Optics Center, the Swiss NSF, and the NCCR QSIT.

\appendix

\section{Alternative way to bosonize}
In this appendix we show that the bosonization of the effective Hamiltonian can also be performed by introducing  bosonic operators for electrons only, $\phi_{r\tau\sigma}$. Thus, introducing bosonic operators for both electrons and holes (`Nambu space representation'), as done in Sec. IV, is not necessary. However, the Nambu space representation is more convenient for time-reversal invariant systems.

In a first step, Eqs. (\ref{H_cN}) - (\ref{H_sN})  become
\begin{align}
{\mathcal H}_{c}^{ee}=&2g_{c}   \Big[\cos(2\phi_{\bar 1 \bar 1 1} +2\phi_{1 1 \bar 1} -\phi_{ 1 \bar 1 1} - \phi_{\bar 1  1 \bar 1} )\nonumber\\
&\hspace{15pt} \cos(2\phi_{\bar 1  1 1} +2 \phi_{1 \bar 1 \bar 1} -\phi_{ 11 1} - \phi_{\bar 1 \bar 1 \bar 1} )\Big],\nonumber\\
{\mathcal H}_{s,\tau}^{ee}=&2g_{\tau} \cos(2\phi_{1 \tau 1}+2{\phi}_{\bar1  \tau \bar 1}-\phi_{\bar 1 \tau1}- {\phi}_{1 \tau \bar 1})\, ,\nonumber\\
\end{align}
leading to
\begin{align}
&{\mathcal H}^{ee}=2g_1  \cos(\eta_{111}+{\eta}_{\bar1 1 \bar 1})+2g_{\bar 1} \cos(\eta_{1 \bar 1 1}+\eta_{\bar 1\bar 1\bar1})\nonumber\\
&\hspace{10pt}+2g_{c} \cos(\eta_{\bar 1 \bar 1 1}+\eta_{1  1 \bar 1})+2g_{c}   \cos(\eta_{ \bar 1  1  1}+\eta_{ 1  \bar 1  \bar 1})
\end{align}
where we introduced  the new chiral fields $\eta_{r\tau\sigma}=2\phi_{r\tau\sigma}-\phi_{\bar r\tau\sigma}$. 
Again, we double the system in order to satisfy the vanishing boundary conditions at the two system ends automatically,
\begin{align}
&\xi_{1\tau}(x)=\begin{cases} \eta_{1 \tau 1}(x), & x>0\\
\eta_{\bar 1  \tau 1}(-x)+\pi, & x<0\, 
\end{cases},\\
&\xi_{\bar 1\tau}(x)=\begin{cases} \eta_{\bar 1 \tau \bar 1}(x), & x>0\\
\eta_{ 1  \tau \bar 1}(-x)+\pi, & x<0\, 
\end{cases}.
\end{align}
Next, we transform the chiral fields to conjugate fields $\phi$ and  $\theta$, via
$\xi_{s\tau} = ( \phi_1+  s \theta_1 +3 \tau \phi_2  +3 \tau s \theta_2)/2$.
As a result, we finally arrive at Eq. (\ref{Nambu_ee}).

\end{document}